\documentclass[twocolumn,pra,superscriptaddress]{revtex4}
\usepackage[colorlinks,linkcolor=blue,anchorcolor=blue,citecolor=blue,filecolor=blue,menucolor=blue,runcolor=blue,urlcolor=blue,frenchlinks=blue]{hyperref}
\usepackage{amsmath}
\usepackage{graphicx}
\usepackage{amssymb}
\usepackage{dcolumn}
\usepackage{mathrsfs}
\usepackage{bm}
\usepackage{color}
\begin{document}

\title{{Proposal of detecting topological transition of quantum braiding in three-fold degenerate eigen subspace}}

\author{Zhi-Wei Han}
\affiliation {Key Laboratory of Atomic and Subatomic Structure and Quantum Control (Ministry of Education), School of Physics, South China Normal University, Guangzhou 510006, China}

\author{Jia-Hao Liang}
\affiliation {Key Laboratory of Atomic and Subatomic Structure and Quantum Control (Ministry of Education), School of Physics, South China Normal University, Guangzhou 510006, China}

\author{Zhao-Xin Fu}
\affiliation {Key Laboratory of Atomic and Subatomic Structure and Quantum Control (Ministry of Education), School of Physics, South China Normal University, Guangzhou 510006, China}

\author{Hong-Zhi Liu}
\affiliation {Key Laboratory of Atomic and Subatomic Structure and Quantum Control (Ministry of Education), School of Physics, South China Normal University, Guangzhou 510006, China}

\author{Zi-Yuan Chen}
\affiliation {Key Laboratory of Atomic and Subatomic Structure and Quantum Control (Ministry of Education), School of Physics, South China Normal University, Guangzhou 510006, China}

\author{Meng Wang}
\affiliation {Key Laboratory of Atomic and Subatomic Structure and Quantum Control (Ministry of Education), School of Physics, South China Normal University, Guangzhou 510006, China}

\author{Ze-Rui He}
\affiliation {Key Laboratory of Atomic and Subatomic Structure and Quantum Control (Ministry of Education), School of Physics, South China Normal University, Guangzhou 510006, China}

\author{Jia-Yi Huang}
\affiliation {Key Laboratory of Atomic and Subatomic Structure and Quantum Control (Ministry of Education), School of Physics, South China Normal University, Guangzhou 510006, China}

\author{Qing-Xian Lv}
\email{LQX0801@163.com}
\affiliation {Key Laboratory of Atomic and Subatomic Structure and Quantum Control (Ministry of Education), School of Physics, South China Normal University, Guangzhou 510006, China}

\affiliation {Guangdong Provincial Key Laboratory of Quantum Engineering and Quantum Materials, South China Normal University, Guangzhou 510006, China}

\affiliation {Guangdong-Hong Kong Joint Laboratory of Quantum Matter, Frontier Research Institute for Physics, South China Normal University, Guangzhou 510006, China}

\author{Kai-Yu Liao}
\email{liaokaiyu1989@163.com}
\affiliation {Key Laboratory of Atomic and Subatomic Structure and Quantum Control (Ministry of Education), School of Physics, South China Normal University, Guangzhou 510006, China}

\affiliation {Guangdong Provincial Key Laboratory of Quantum Engineering and Quantum Materials, South China Normal University, Guangzhou 510006, China}

\affiliation {Guangdong-Hong Kong Joint Laboratory of Quantum Matter, Frontier Research Institute for Physics, South China Normal University, Guangzhou 510006, China}

\author{Yan-Xiong Du}
\email{yanxiongdu@m.scnu.edu.cn}
\affiliation {Key Laboratory of Atomic and Subatomic Structure and Quantum Control (Ministry of Education), School of Physics, South China Normal University, Guangzhou 510006, China}

\affiliation {Guangdong Provincial Key Laboratory of Quantum Engineering and Quantum Materials, South China Normal University, Guangzhou 510006, China}

\affiliation {Guangdong-Hong Kong Joint Laboratory of Quantum Matter, Frontier Research Institute for Physics, South China Normal University, Guangzhou 510006, China}

%\date{\today}

\begin{abstract}
The braiding operations of quantum states have attracted substantial attention due to their great potential for realizing topological quantum computations. In this paper, we show that a three-fold degenerate eigen subspace can be obtained in a four-level Hamiltonian which is the minimal physical system. Braiding operations are proposed to apply to dressed states in the subspace.  The topology of the braiding diagram can be characterized through physical methods once that the sequential braiding pulses are adopted. We establish an equivalent relationship function between the permutation group and the output states where different output states correspond to different values of the function. The topological transition of the braiding happens when two operations overlap, which is detectable through the measurement of the function. Combined with the phase variation method, we can analyze the wringing pattern of the braiding. Therefore, the experimentally-feasible system provides a platform to investigate braiding dynamics, the SU(3) physics and the qutrit gates.
\end{abstract}

 \maketitle

\section{Introduction}
Topological quantum computation has been recognized as one of the most important approaches toward realizing the fault-tolerant quantum computer \cite{Nayak2008,kiteav2003,Fredman2002}. The scheme relies on the non-Abelian anyons which exist in the degenerate eigen subspace and obey non-Abelian braiding statistics. The unitary gate operations that used to realize quantum computation are carried out by braiding non-Abelian anyons and measuring the final states. The fault-tolerance of the topological quantum computer arises from the nonlocal encoding of the qubits, which makes them immune to errors caused by local perturbations. There have been numerous physical systems proposed to realize topological quantum computers, i.e., the fractional-Hall states \cite{Moore1991,Bonderson2006}, cold atoms \cite{Deng2015}, topological superconductors \cite{Ivanov2001}, and the Majorana zero-modes \cite{Alicea2011}. However, substantial challenges still exist in the real experimental realizations. Recently, two research groups independently realized the simulation of the Ising-type non-Abelian aynons (also called Ising aynons) in superconducting quantum simulators with serial digital quantum gates \cite{google2023,Xu2023}. The non-Abelian feature and the fusion rules are experimentally simulated in such systems. The above works have demonstrated the topological quantum computation to be experimentally feasible.

On the other hand, analog experiments of non-Abelian braiding operations which connect to the topological quantum gates are also investigated \cite{Zhang2022,Scheel2022,Chen2022,Chen2019,Guo2021,Neef2023,Wojcik2020,Li2021,Yang2020,Hu2021,Hu2022}. To reveal the non-Abelian characteristic, three-fold degenerate eigen states are needed which can be constructed by single-photon interacting with seven waveguide modes in photonic chips \cite{Zhang2022}. Different light-diffraction patterns correspond to different braiding orders. It is intrinsical the geometric effects since the dynamical phases are trivial during the evolution process. Similar results are also performed in acoustic waveguide modes \cite{Chen2022}. Furthermore, the three-fold degenerate states can be induced with two photons interacting with four waveguide modes in photonic chips, which is used to realize three-dimensional quantum holonomy \cite{Neef2023}. At the same time, it seems that the existed braiding results have no difference from mathematical results with classical braiding (that is, braiding classical objects with different orders will obtain different final states). It would be interesting to ask whether a `quantum' braiding (braiding quantum systems with quantum controls) would be different from a `classical' braiding. To answer the above question, one may resort to a quantum system with a fully controllable Hamiltonian.

In this paper, we propose to perform non-Abelian braiding operations in a three-fold degenerate subspace in cold atomic system. The degenerate eigen subspace is constructed in a four-level system which costs the minimal physical resources. Braiding operations of the eigen states are introduced in such systems by imposing dressed pulses upon dressed states. Different braiding orders generate different final states which shows the non-Abelian character. The quantumness of the braiding operations is shown by overlapping two operations that will make topological transition happen. To quantify such transitions, mapping function that are experimentally measurable are introduced to label different final output states. Therefore, the proposal system with three-fold degenerate subspace provides an experimentally-feasible way to investigate the braiding dynamics, and furthermore, the SU(3) physics \cite{Barnett2012,Hu2014}. The paper is organized as follows: In section II we introduce the four-level system that generates the three-fold degenerate eigen subspace. In section III we introduce the braiding operation upon dressed states and test the robustness. In section IV we character the topology of the linking pattern by introducing an equivalence relationship which is experimentally detectable. In section V we character the topology of the wringing pattern by testing the response to the phase variation of $\pi$ pulses. We conclude the paper in section VI.

%%%%%%%%%%%%%%%%%%%%%%%%%%%%%%%%%%%%%%%%%%%%%%%%%%%
\begin{figure}[ptb]
\begin{center}
\includegraphics[width=8.5cm]{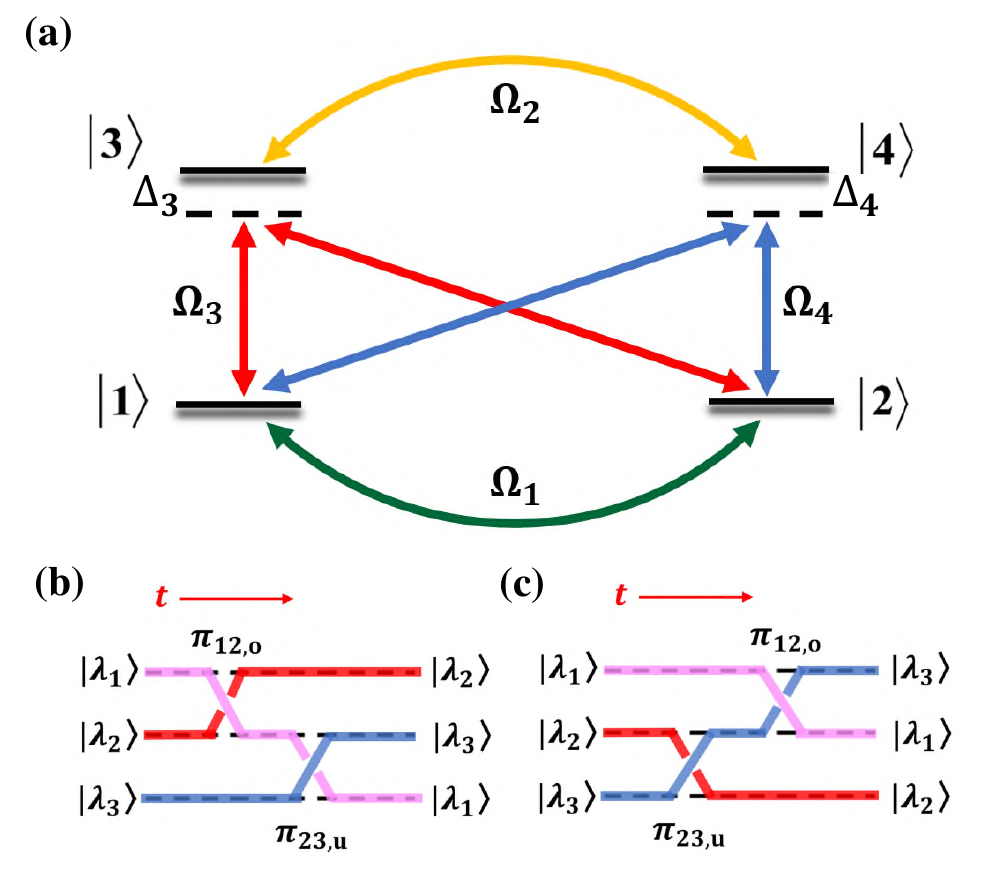}
\caption{\label{fig:level} (a) Coupling scheme that used to achieve the three-fold degenerate eigen subspace. Bare states $\{|1\rangle, |2\rangle, |3\rangle, |4\rangle\}$ are four energy levels in a cold $^{87}$Rb atomic system. $\Omega_1, \Omega_2$ are the Rabi frequencies of Radio fields or two-photon Raman transitions while $\Omega_3(\Omega_4)$ with detuning $\Delta_3(\Delta_4)$ are the Rabi frequencies of microwave fields.  (b) World-line description of braiding schemes of three degenerate eigen states $\{|\lambda_1\rangle, |\lambda_2\rangle, |\lambda_3\rangle\}$. $t$ denote the evolving time and the solid colored lines symbol the positions of the states. Different orders of braiding operations $\{\pi_{12,\mathrm{o}}, \pi_{23,\mathrm{u}}\}$ will induce different final states which confirms the non-Abelian characteristic of the braiding. The orientation of the braiding can be introduced by introducing the rotating direction of the $\pi$ pulses where the over-crossing operation is achieved by $\pi_{kj,\mathrm{o}}$ in Eq.(5) and the under-crossing is achieved by $\pi_{kj,\mathrm{u}}$ in Eq.(6).}
\end{center}
\end{figure}

%%%%%%%%%%%%%%%%%%%%%%%%%%%%%%%%%%%%%%%%%%%%%%%%%

\section{Hamiltonian with three-fold degenerate eigen subspace}

%%%%%%%%%%%%%%%%%%%%%%%%%%%%%%%%%%%%%%%%%%%%%%%%%%%
\begin{figure*}[ptb]
\begin{center}
\includegraphics[width=16.0cm]{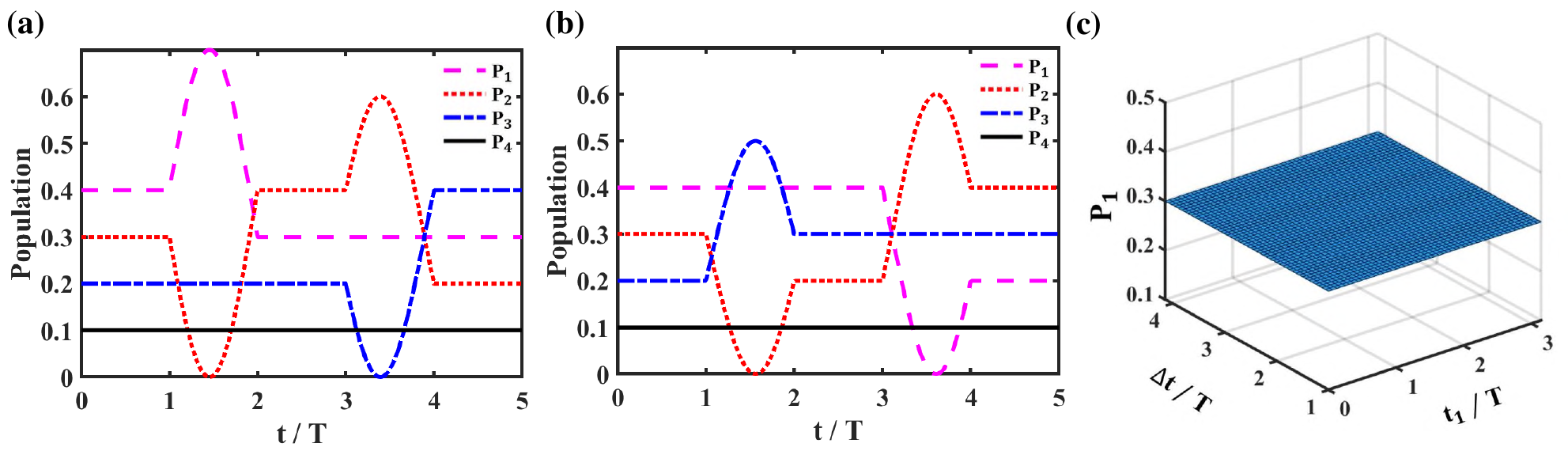}
\caption{The braiding dynamics and their robustness. (a) The braiding dynamics of union operation $\pi_{23,\mathrm{u}}\pi_{12,\mathrm{o}}$. (b) The braiding dynamics of union operation $\pi_{12,\mathrm{o}}\pi_{23,\mathrm{u}}$. The initial states of both cases are set to be $|\Psi_0\rangle=\sqrt{0.4}|\lambda_1\rangle+\sqrt{0.3}|\lambda_2\rangle+\sqrt{0.2}|\lambda_3\rangle+\sqrt{0.1}|\lambda_4\rangle$. Hamiltonian (1) with $\theta=\pi/2$, $\alpha=\pi/2$, and $\varphi=0$ is introduced between the gap of the braiding pulses which makes the system always evolve in the eigen subspace. The evolution is thus governed by $U=U_H(T_3)U_\pi(T)U_H(T_2)U_\pi(T)U_H(T_1)$, $U_\pi(T)=\{\pi_{12,\mathrm{o}}, \pi_{23,\mathrm{u}}\}$,  $T_1=T_2=T_3=T$, $T$ is the evolution period of the braiding pulses. Pink dashed lines: population of $|\lambda_1\rangle$, $P_1$; Red dotted lines: population of $|\lambda_2\rangle$, $P_2$; Blue dashed-dotted lines: population of $|\lambda_3\rangle$, $P_3$; Black solid lines: population of $|\lambda_4\rangle$, $P_4$; The population of eigen states can be detected by the ones of bare states with suitable pulses. (c) Population $P_1$ versus starting time $t_1$ and the time interval $\Delta t$.}
\end{center}
\end{figure*}
%%%%%%%%%%%%%%%%%%%%%%%%%%%%%%%%%%%%%%%%%%%%%%%%%%%

Here we construct a three-fold degenerate eigen subspace in a fully controllable cold atomic system.  We adopt a four-level system interacting with six optical fields as shown in Fig. 1(a), where the energy levels are chosen to be $|1\rangle=|F=2,m_F=-1\rangle$, $|2\rangle=|F=1,m_F=-1\rangle$, $|3\rangle=|F=2,m_F=0\rangle$ and $|4\rangle=|F=1,m_F=0\rangle$ in $^{87}$Rb atom. Under the bare state basis $\{|1\rangle, |2\rangle, |3\rangle, |4\rangle\}$ and rotating-wave approximation, the interaction Hamiltonian is given by
\begin{equation}
H=\frac{1}{2}\left ( \begin{array}{cccc}
0&\Omega_1 e^{i\varphi}&\Omega_3& \Omega_4 e^{i\varphi}\\
\Omega_1 e^{-i\varphi}& 0 & \Omega_3 e^{-i\varphi} &\Omega_4\\
\Omega_3&\Omega_3 e^{i\varphi}& \Delta_3 &\Omega_2 e^{i\varphi}\\
\Omega_4 e^{-i\varphi}&\Omega_4&\Omega_2 e^{-i\varphi} &  \Delta_4
\end{array}\right),
\end{equation}
here we have adopted $\hbar=1$. The Rabi frequencies $\Omega_3(\Omega_4)$ with detuning $\Delta_3(\Delta_4)$ that coupling $\{|1\rangle, |3\rangle\}, \{|2\rangle, |3\rangle\}(\{|1\rangle, |4\rangle\}, \{|2\rangle, |4\rangle\})$ are set to be microwave fields, where similar experiments have realized in \cite{Lv2021}. The coupling $\Omega_1(\Omega_2)$ between $\{|1\rangle, |2\rangle\}(\{|3\rangle, |4\rangle\})$ can be realized by radio fields \cite{Sugawa2018,Sugawa2021} or two-photon Raman transitions \cite{Lin2011}. One can find that when the detuning satisfying conditions $\Delta_3=\Omega_3^2/\Omega_1-\Omega_1, \Delta_4=\Omega_4^2/\Omega_1-\Omega_1, \Omega_1\Omega_2=\Omega_3\Omega_4$, the eigenvalues of  Hamiltonian (1) are given by
\begin{equation}
\lambda_1=\lambda_2=\lambda_3=-\Omega_1, \lambda_4=\frac{\Omega_1^2+\Omega_2^2+\Omega_3^2}{\Omega_1^2},
\end{equation}
of which a three-fold degenerate subspace with the lowest eigenvalues exist. Comparing with the seven-level scheme \cite{Zhang2022} or the $N-$pod system \cite{Leroux2018,Zhang2018}, the proposed system is the minimal physical one, promoting the experimental feasibility. By parameterizing $\Omega_1=\Omega_0\sin\alpha\sin\theta, \Omega_2=2\Omega_0\cos\alpha/\tan\theta, \Omega_3=\sqrt{2}\Omega_0\sin\alpha\cos\theta, \Omega_4=\sqrt{2}\Omega_0\cos\alpha$, the eigen states will be obtained as
\begin{equation}
\begin{split}
&|\lambda_1\rangle=\frac{\sqrt{2}}{2}(|1\rangle-e^{-i\varphi}|2\rangle),\\
&|\lambda_2\rangle=\cos\theta|b\rangle-\sin\theta|3\rangle,\\
&|\lambda_3\rangle=\cos\alpha|c\rangle-e^{-i\varphi}\sin\alpha|4\rangle,\\
&|\lambda_4\rangle=\sin\alpha|c\rangle+e^{-i\varphi}\cos\alpha|4\rangle,
\end{split}
\end{equation}
where $|b\rangle=\sqrt{2}/2(|1\rangle+e^{-i\varphi}|2\rangle), |c\rangle=\sin\theta|b\rangle+\cos\theta|3\rangle$.
Under the subspace spanned by the lowest eigenstates, one will obtain the non-Abelian gauge potentials and gauge fields as calculated by $A^{jk}_\mu=i\langle \lambda_j|\partial_\mu|\lambda_k\rangle$ ($A^{jk}_\mu$ is the matrix elements of $A_{lm, \mu}$ which is defined in the two of three-fold eigen states, i.e., the subspace of $\{|\lambda_1\rangle, |\lambda_2\rangle\}$) and $F_{lm, \mu\nu}=\partial_\mu A_{lm, \nu}-\partial_\nu A_{lm, \mu}-i[A_{lm, \mu}, A_{lm, \nu}]$  , respectively \cite{Wilzeck1984,Duan2001}. Using Eq.(3), the non-diagonal matrix of SU(3)  gauge field $F_{lm,\theta\alpha}$ will be given by
\begin{eqnarray}\nonumber
&&F_{12,\theta\alpha}=F_{13,\theta\alpha}=0,\\
&&F_{23,\theta\alpha}=i\left ( \begin{array}{cc}
0& \dot{\theta}\dot{\alpha}\sin\alpha\\
-\dot{\theta}\dot{\alpha}\sin\alpha & 0
\end{array}\right).
\end{eqnarray}
Therefore, the three-fold degenerate eigen subspace provide a fully controllable platform to investigate the braiding dynamics and the evolution will be determined by geometric characteristic of the gauge field which can be treated as geometric quantum control \cite{Sj2012,Abdumalikov,Zu2014,Toyoda2013,Zhu2002,Xue201802,Xue201801,Xue2021,Xue2023,Du2017}.

\section{Braiding operation of dressed states and the robustness}
In the following, we discuss the braiding operations in the eigen subspace constructed by $\{|\lambda_1\rangle, |\lambda_2\rangle, |\lambda_3\rangle\}$. Considering the oriented link of the braiding configuration \cite{Adams1995,Yang1994,Freivalds2005}, the over-crossing braiding operation can be realized by $\pi$ pulses upon dressed states which is defined as
\begin{equation}
\pi_{kj,\mathrm{o}}=-i(|\lambda_k\rangle\langle\lambda_j|e^{i\phi_{kj}}+|\lambda_j\rangle\langle\lambda_k|e^{-i\phi_{kj}})+|\lambda_l\rangle\langle\lambda_l|+|\lambda_4\rangle\langle\lambda_4|,
\end{equation}
where $k, j, l=1, 2, 3$. And the under-crossing braiding is defined as
\begin{equation}
\pi_{kj,\mathrm{u}}=i(|\lambda_k\rangle\langle\lambda_j|e^{i\phi_{kj}}+|\lambda_j\rangle\langle\lambda_k|e^{-i\phi_{kj}})+|\lambda_l\rangle\langle\lambda_l|+|\lambda_4\rangle\langle\lambda_4|,
\end{equation}
which satisfies the relationships $\pi_{kj,\mathrm{o}}\pi_{kj,\mathrm{o}}=\pi_{kj,\mathrm{u}}\pi_{kj,\mathrm{u}}=-1, \pi_{kj,\mathrm{o}}\pi_{kj,\mathrm{u}}=\pi_{kj,\mathrm{u}}\pi_{kj,\mathrm{o}}=1$ \cite{Note0}. It can be seen that the orientation of the crossing is equivalent to the rotating direction of the quantum states along a certain axis. Once the mapping between the braiding and the $\pi$ pulses sequences is established, we can investigate the topology of the braiding in the dressed states subspace.

The $\pi$ pulses $\pi_{kj,\mathrm{o}}$ is connected to the evolution governed by the Hamiltonian $H_{kj}$ with $\pi_{kj,\mathrm{o}}=e^{-iH_{kj}T}$ and $\pi_{kj,\mathrm{u}}=e^{iH_{kj}T}$, $T$ is the evolution period. To simplify the experimental realization, we set the initial control parameters with $\theta(t=0)=\pi/2, \alpha(t=0)=\pi/2, \varphi(t=0)=0$ \cite{Note1} where the eigen states turn out to be
\begin{equation}
\begin{split}
&|\lambda_1\rangle_{in}=\frac{\sqrt{2}}{2}(|1\rangle-|2\rangle), |\lambda_2\rangle_{in}=-|3\rangle,\\
&|\lambda_3\rangle_{in}=-|4\rangle,|\lambda_4\rangle_{in}=\frac{\sqrt{2}}{2}(|1\rangle+|2\rangle).
\end{split}
\end{equation}
In this case, we find that the braiding operations $\pi_{12,\mathrm{0}}$,$\pi_{23,\mathrm{u}}$ with $\phi_{12}=\phi_{23}=\phi$ can be realized by the Hamiltonian under the bare states basis $\{|1\rangle, |2\rangle, |3\rangle, |4\rangle\}$ with
\begin{equation}
H_{12}=\frac{\Omega_{12}}{4}\left ( \begin{array}{cccc}
0&0&-\sqrt{2}e^{i\phi}& 0\\
0& 0 & \sqrt{2}e^{i\phi} &0\\
-\sqrt{2}e^{-i\phi}&\sqrt{2}e^{-i\phi}& 0 &0\\
0&0&0 & 0
\end{array}\right),
\end{equation}
\begin{equation}
H_{23}=\frac{\Omega_{23}}{2}\left ( \begin{array}{cccc}
0&0&0& 0\\
0& 0 & 0&0\\
0&0& 0 &e^{i\phi}\\
0&0&e^{-i\phi} &  0
\end{array}\right),
\end{equation}
with $\Omega_{12} T=\Omega_{23} T=\pi$. It can be seen that $H_{12}$ can be realized by a triangle-type three-level system \cite{Du2017} while $H_{23}$ is achieved by coupling the two-level system $\{|3\rangle, |4\rangle\}$.

As shown in Fig. 1(b) and 1(c), braiding operations of groups $\{\pi_{12,\mathrm{o}} , \pi_{23,\mathrm{u}}\}$ are introduced, where $\pi_{12,\mathrm{o}}$ is used to inverse the population of dressed states $|\lambda_1\rangle$ and $|\lambda_2\rangle$, and $\pi_{23,\mathrm{u}}$ is used to inverse the population of dressed states $|\lambda_2\rangle$ and $|\lambda_3\rangle$. Mathematically, the final braiding states are determined by the orders of braiding operations which reveal the non-Abelian characteristic. For example, when the systems are prepared in the state $|\Psi_0\rangle=a_1|\lambda_1\rangle+a_2|\lambda_2\rangle+a_3|\lambda_3\rangle+a_4|\lambda_4\rangle$, the final state will be $|\Psi_1\rangle=a_2|\lambda_1\rangle+a_3|\lambda_2\rangle+a_1|\lambda_3\rangle+a_4|\lambda_4\rangle$ against the union operations $\pi_{23,\mathrm{u}}\pi_{12,\mathrm{o}}$ (Fig. 1(b)), while $|\Psi_2\rangle=a_3|\lambda_1\rangle+a_1|\lambda_2\rangle+a_2|\lambda_3\rangle+a_4|\lambda_4\rangle$ correspond to the one of $\pi_{12,\mathrm{o}}\pi_{23,\mathrm{u}}$ (Fig. 1(c)). To detect the final output state, one needs to measure the population on each eigen states which is done by applying a $\pi/2$-pulse to $\{|1\rangle, |2\rangle\}$. Therefore, the correspondence between the final braiding states and the bare states will be given by $|\lambda_1\rangle_f=|1\rangle, |\lambda_2\rangle_f=-|3\rangle, |\lambda_3\rangle_f=-|4\rangle, |\lambda_4\rangle_f=|2\rangle$ and the eigen states can be detected by measuring the bare states.

In Fig. 2(a) and 2(b), the population dynamics of the braiding upon dressed states representations are depicted. The initial state is chosen to be $|\Psi_0\rangle=\sqrt{0.4}|\lambda_1\rangle+\sqrt{0.3}|\lambda_2\rangle+\sqrt{0.2}|\lambda_3\rangle+\sqrt{0.1}|\lambda_4\rangle$,
and Fig. 2(a) are the results of operations $\pi_{23,\mathrm{u}}\pi_{12,\mathrm{o}}$ while Fig. 2(b) are the ones of operations $\pi_{12,\mathrm{o}}\pi_{23,\mathrm{u}}$, with $\phi=\pi/2$. To investigate the general dynamics, Hamiltonian (1) with $\theta=\pi/2$, $\alpha=\pi/2$, and $\varphi=0$ is applied between the gap of the braiding pulses which makes the system always evolve in the eigen subspace \cite{Note2}, that is, the evolution is given by
\begin{equation}
U=U_H(T_3)U_\pi(T)U_H(T_2)U_\pi(T)U_H(T_1),
\end{equation}
$U_H(T_i)=e^{-iHT_i}, i=1, 2, 3$, $T_i$ are the interaction duration of $H$, $U_\pi(T)=\{\pi_{23,\mathrm{u}}, \pi_{12,\mathrm{o}}\}$. In the numerical simulation, we set $\Omega_{12}=\Omega_{23}=1$ and $T_1=T_2=T_3=T=\pi$.  $U_H$ will not change the population upon the eigen states since $|\lambda_i\rangle_{in}, i=1, 2, 3$ are the instantaneous eigen states of $H$. As shown in Fig. 2(a) and 2(b), the different braiding results of different braiding orders truly reveal the non-Abelian characteristic.

Here we discuss the robustness of braiding, as shown in Fig. 2(c). Assuming the starting time of $\pi_{12,\mathrm{o}}(\pi_{23,\mathrm{u}})$ is set to be $t_1 (t_2)$,  we tilt $t_1 (t_2)$ and observe the final population of union operations $\pi_{23,\mathrm{u}}\pi_{12,\mathrm{o}}$. It can be seen that $t_1=T_1, t_2=T_1+T+T_2$. By keeping the separation $\Delta t=t_2-t_1>T$ (two braiding operations do not overlap), the population $P_1$ of eigen state $|\lambda_1\rangle$ do not change along with $t_1$ and $\Delta t$ (other populations are also similar with $P_1$). Such robustness is guaranteed by the topology of the braiding which does not change as long as one changes the control parameters.

%%%%%%%%%%%%%%%%%%%%%%%%%%%%%%%%%%%%%%%%%%%%%%%%%
\begin{figure}[ptb]
\begin{center}
\includegraphics[width=8.5cm]{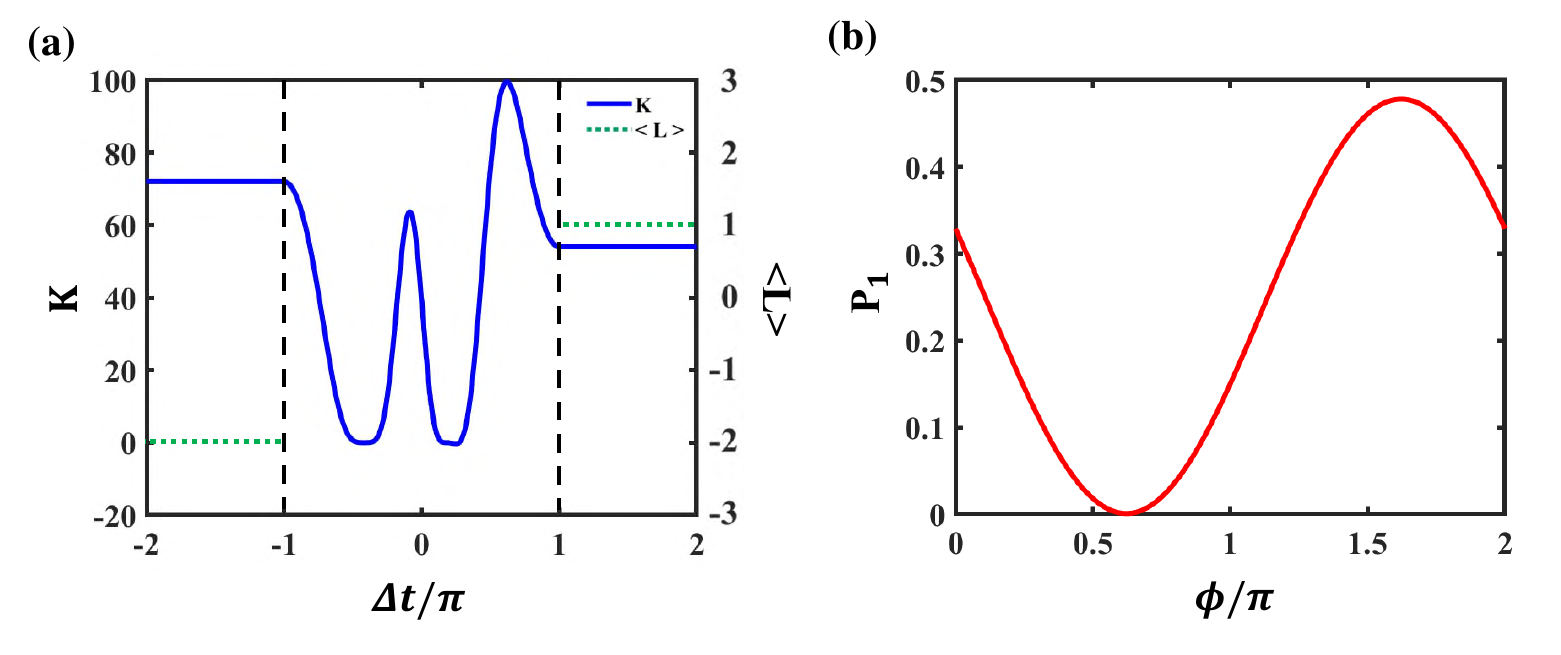}
\caption{(a) Equivalent relationship function $K$ of braiding $\{\pi_{23,\mathrm{u}}, \pi_{12,\mathrm{o}}\}$ versus the deviation of $\Delta t$. $\Delta t>\pi$ and $\Delta t<-\pi$ correspond to two different braiding configurations and $K$ will change accordingly. The classification of topology by $K$ is the same as the bracket polynomial $<L>$ in the well-defined regions. The topological transition will happen when two braiding pulses overlap of which the dynamics can be described by $K$. (b) Relationship between population $P_1$ and the relative phase $\phi$. The phase coherence shows the quantumness of the braiding. }
\end{center}
\end{figure}
%%%%%%%%%%%%%%%%%%%%%%%%%%%%%%%%%%%%%%%%%%%%%%%%%%%

\section{Characterizing the linking pattern and topological transition}
In the following, we discuss how to character the braiding configuration with physical methods in the dressed state subspace, and make a comparison with the one in mathematic theory.

We first briefly review how to character the topology of braiding in mathematic theory \cite{Adams1995,Yang1994,Freivalds2005,Collins2006}. Since the braiding diagrams can be changed to knots by certain rules, we use the language of knot theory in the following. To determine the topology, one needs to quantify the linking patterns (the numbers and the distribution of the crossing) and wringing patterns (the orientation of the crossing) which utilize the Jones polynomial. Given an oriented link $L$, the Jones polynomial is given by $X(L)=(-A^3)^{-\omega(L)}<L>$, where $A$ is a variable, $<L>$ is the bracket polynomial that character the linking patterns of $L$ and $\omega(L)$ is called writhe that character the wringing patterns of $L$. It can be found that the topological invariant in knot theory is different from the Chern number in condensed matter physics which is defined as an integral of gauge field (described by the mathematics called differential geometry) \cite{Lv2021,Schroer2014,Roushan2014,Sugawa2018,XTan2021}.

In the following, we will introduce how to character the topology of braiding in dressed state subspace with physical methods.

We first discuss the characterization of the linking pattern of the braiding. Assuming that there are $m$ input states with the population denoted as $P_i, i=1, 2,..., m$ with $P_1>P_2>\cdots>P_m$. We can redefine $P_i$ by $P'_i=m+1-i$. One can find that the final output state after braiding will be one of the permutations of arrangement $(P'_1P'_2...P'_m)$ which can be marked as
\begin{equation}
\xi=\left ( \begin{array}{cccc}
P'_1&P'_2&\cdots&P'_m\\
A_1& A_2 & \cdots&A_m\\
\end{array}\right).
\end{equation}
Then we map the permutation $\xi$ to an one to one function as
\begin{equation}
K(\xi)=\prod_i (A_i)^i.
\end{equation}
$K$ is experimentally determined from the population $P_i$. It is known that the permutation groups are the invariant subgroup of braiding groups \cite{Yang1994} and the output result will not be changed by the three Reidemeister moves in knot theory, function $K$ can be treated as an equivalence relationship which separate the braiding into several groups. Since the braiding scheme with $N$ crossing will have $2^N$ possible kinds of configuration and $m$ states will have $m!$ possible permutations, we can construct a physical system with big enough $m$ to cover the possible configurations of the braiding.

In Fig. 3(a), index $K$ versus separation time $\Delta t$ of two braiding operations $\{\pi_{23,\mathrm{u}}, \pi_{12,\mathrm{o}}\}$ are depicted for the case of $m=3$. The initial state are chosen to be $|\Psi_0\rangle$ (the same as in Fig. 2). The fully controllable quantum system offers a platform to investigate the topological transition dynamics of braiding. When $\Delta t>\pi$, two braiding operations are well separated, $K$ is stable at 54 as the given final state $|\Psi_1\rangle=\sqrt{0.3}|\lambda_1\rangle+\sqrt{0.2}|\lambda_2\rangle+\sqrt{0.4}|\lambda_3\rangle+\sqrt{0.1}|\lambda_4\rangle$ after the union operations $\pi_{23,\mathrm{u}}\pi_{12,\mathrm{o}}$. $K$ will be stable at 72 when $\Delta t<-\pi$, where final state $|\Psi_2\rangle=\sqrt{0.2}|\lambda_1\rangle+\sqrt{0.4}|\lambda_2\rangle+\sqrt{0.3}|\lambda_3\rangle+\sqrt{0.1}|\lambda_4\rangle$ after the union operation $\pi_{12,\mathrm{o}}\pi_{23,\mathrm{u}}$. As can be seen that the output states are robust against the separation time between pulses if the pulses are well separated. The above classification results are in accord with the ones of the bracket polynomial $<L>$. To calculate $<L>$, we insert the braiding $\{\pi_{23,\mathrm{u}}, \pi_{12,\mathrm{o}}\}$ in a non-trivial knot which induce $<L>=1$ for the case $\pi_{23,\mathrm{u}}\pi_{12,\mathrm{o}}$ and $<L>=-2$ for the case $\pi_{12,\mathrm{o}}\pi_{23,\mathrm{u}}$.

When $-\pi<\Delta t<\pi$, braiding operations overlap. Here $K$ is sensitive to $\Delta t$ since the overlapping states are not topological. Therefore, we can detect the dynamics of topological transition in the dressed states subspace which cannot be realized in classical systems.
%%%%%%%%%%%%%%%%%%%%%%%%%%%%%%%%%%%%%%%%%%%%%%%%%
\begin{figure}[ptb]
\begin{center}
\includegraphics[width=8.5cm]{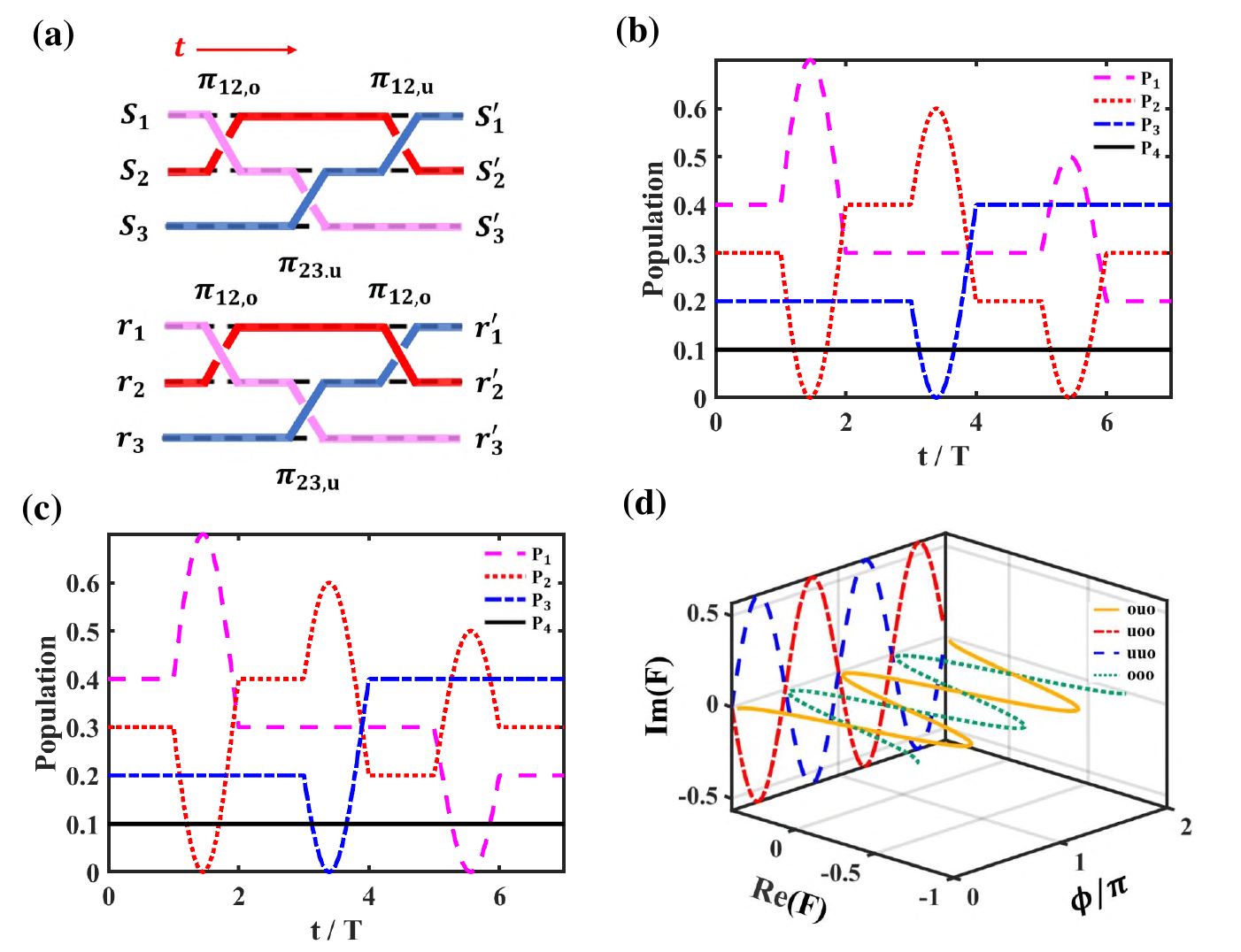}
\caption{(a) World-line description of braiding $\pi_{12,\mathrm{u}}\pi_{23,\mathrm{u}}\pi_{12,\mathrm{o}}$ (upper panel) and $\pi_{12,\mathrm{o}}\pi_{23,\mathrm{u}}\pi_{12,\mathrm{o}}$ (bottom panel) where only the orientation of the final $\pi$ pulses of the two braiding are different. (b) Population dynamics against braiding $\pi_{12,\mathrm{u}}\pi_{23,\mathrm{u}}\pi_{12,\mathrm{o}}$. (c) Population dynamics against braiding $\pi_{12,\mathrm{o}}\pi_{23,\mathrm{u}}\pi_{12,\mathrm{o}}$. Pink dashed lines: population of $|\lambda_1\rangle$, $P_1$; Red dotted lines: population of $|\lambda_2\rangle$, $P_2$; Blue dashed-dotted lines: population of $|\lambda_3\rangle$, $P_3$; Black solid lines: population of $|\lambda_4\rangle$, $P_4$. (d) Inner product $F=\langle\Psi_f|\Psi_i\rangle$ versus the phase variation of the $\pi$ pulses, $|\Psi_i\rangle$: the initial state, $|\Psi_f\rangle$: the final state, $\mathrm{Re}(F)$: the real part of $F$, $\mathrm{Im}(F)$: the imaginary part of $F$. Yellow-solid line: braiding of $\pi_{12,\mathrm{o}}\pi_{23,\mathrm{u}}\pi_{12,\mathrm{o}}$; Red dashed-dotted line: braiding of $\pi_{12,\mathrm{u}}\pi_{23,\mathrm{o}}\pi_{12,\mathrm{o}}$; Blue-dashed line: braiding of $\pi_{12,\mathrm{u}}\pi_{23,\mathrm{u}}\pi_{12,\mathrm{o}}$; Green-dotted line: braiding of $\pi_{12,\mathrm{o}}\pi_{23,\mathrm{o}}\pi_{12,\mathrm{o}}$. All the calculations use the same initial state $|\Psi_0\rangle$ as in Fig. 2.}
\end{center}
\end{figure}
%%%%%%%%%%%%%%%%%%%%%%%%%%%%%%%%%%%%%%%%%%%%%%%%%%%

\section{Characterizing the wringing pattern with phase variation}
Here we discuss the characterization of the wringing pattern of the braiding. As for the oriented link, the linking pattern is not enough to determine the topology of the braiding. Different from the writhe $\omega(L)$ in Jones polynomial, here we propose to investigate the wringing patterns with the phase dynamics of the quantum systems. To test the phase coherence of the operations $\pi_{23,\mathrm{u}}\pi_{12,\mathrm{o}}$, we tilt the relative phases between the two braiding operations and calculate the final states, as described in Fig. 3(b). We set $\phi_{12}=0$ and change $\phi$ from 0 to $2\pi$ in $H_{23}$. The final population of $P_1$ will change along with $\phi$. Similar cases happen in $P_2$ and $P_3$ which will not happen in classical braiding. Therefore, the dynamics of braiding in quantum systems will be more interesting and may lead to novel phenomena.

In Fig. 4, we discuss the braiding with three braiding pulses by two cases: $\pi_{12,\mathrm{u}}\pi_{23,\mathrm{u}}\pi_{12,\mathrm{o}}$ (upper panel in Fig. 4(a)) and $\pi_{12,\mathrm{o}}\pi_{23,\mathrm{u}}\pi_{12,\mathrm{o}}$ (bottom panel in Fig. 4(a)), where the only difference between the two braiding is the orientation of the final pulses. Such braiding schemes are frequently studied in the investigation of Yang-Baxter equations \cite{Zheng2013,AnvariVind2016,1Wang2020}. The population dynamics in Fig. 4(b) corresponds to the braiding $\pi_{12,\mathrm{u}}\pi_{23,\mathrm{u}}\pi_{12,\mathrm{o}}$ and the ones in Fig. 4(c) corresponds to the braiding $\pi_{12,\mathrm{o}}\pi_{23,\mathrm{u}}\pi_{12,\mathrm{o}}$, where the input state $|\Psi_0\rangle$ is the same as Fig. 2.  Since the population of the output states for both of the braiding are the same, function $K$ can not distinguish the two braiding. The problem also exists in the bracket polynomial $<L>$ in knot theory. By connecting the starting point and the ending point that lay in the same dashed lines in the world-lines description of the braiding (i.e., $s_1$ and $s'_1$, $s_2$ and $s'_2$ and $s_3$ and $s'_3$ in Fig. 4(a)), we will obtain knots or chains from the braiding diagram in Fig. 4(a), where both braiding diagram give the same value of $<L>=-2$. To further character the topology, we calculate the final states $|\Psi_f\rangle$ against the variation of the phase $\phi_{kj}$ in the $\pi$ pulses of Eq.(5) and Eq.(6). In Fig. 4(d), we plot the results of $F=\langle\Psi_f|\Psi_i\rangle$ against $\phi$ of which the y-axis symbols the values of the real parts $\mathrm{Re}(F)$ and z-axis symbols the values of the imaginary parts $\mathrm{Im}(F)$ of $F$. As can be seen, the response curve of braiding $\pi_{12,\mathrm{u}}\pi_{23,\mathrm{u}}\pi_{12,\mathrm{o}}$ (blue-dashed line in Fig. 4(d)) is different from the one of $\pi_{12,\mathrm{o}}\pi_{23,\mathrm{u}}\pi_{12,\mathrm{o}}$ (yellow-solid line in Fig. 4(d)). Another cases of braiding $\pi_{12,\mathrm{u}}\pi_{23,\mathrm{o}}\pi_{12,\mathrm{o}}$ (red dashed-dotted line) and $\pi_{12,\mathrm{o}}\pi_{23,\mathrm{o}}\pi_{12,\mathrm{o}}$ (green-dotted line) are also plotted where the response curves against the phase variation are all different.  The above classification results are similar to the ones of using writhe $\omega$ in Jones polynomial. By defining the over-crossing to be positive crossing and the under-crossing to be the negative crossing, the writhe (number of positive crossing minus the number of negative crossing) will also derive four possible values $\pm 3$ and $\pm 1$ \cite{Note3}. Therefore, the phase coherence of the quantum system offers a physical method to investigate the topology of the braiding.

\section{Scalability of the proposed system}
Here we discuss the scalability of the proposed system. According to the above discussion, one can achieve a $(N-1)$-fold degenerate eigen subspace from a $N$-level system. Considering the present experimental system, the maximal controllable levels will be $N_{\mathrm{max}}=128$ with the hyperfine ground states in Holmium atom \cite{Saffman2008,Hostetter2015}. To achieve a larger Hilbert space, one may adopt the multi-particle states of the interacting system instead of the atomic levels, i.e., $N$ interacting two-level atoms correspond to $2^N$ possible states. Assuming that an input state is given by $|\Psi_0\rangle=\sum_i c_i|\xi_i\rangle$, where $|\xi_i\rangle$ is one of the possible multi-particle states $|\xi_i\rangle=|\varepsilon_1\rangle\otimes|\varepsilon_2\rangle\cdots\otimes|\varepsilon_N\rangle$, $|\varepsilon_k\rangle, k=1,\cdots,N,$ are the quantum states of the $k$th two-level atom which is labelled by $\{|0\rangle, |1\rangle\}$. The braiding operation between $|\xi_k\rangle$ and $|\xi_j\rangle$ is given by
\begin{equation}
\pi'_{kj,\mathrm{o(u)}}=\mp i(|\xi_k\rangle\langle\xi_j|e^{i\phi_{kj}}+|\xi_j\rangle\langle\xi_k|e^{-i\phi_{kj}})+\sum_{l\neq k,j}|\xi_l\rangle\langle\xi_l|,
\end{equation}
which can be realized by the evolution $\pi'_{kj,\mathrm{o(u)}}=e^{\mp i\int_0^{T_I} H'_{kj}t}$ governed by the Hamiltonian $H'_{kj}=\frac{\Omega_{kj}}{2}(|\xi_k\rangle\langle\xi_j|e^{i\phi_{kj}}+|\xi_j\rangle\langle\xi_k|e^{-i\phi_{kj}})$, with $\Omega_{kj}T_I=\pi$. Hamiltonian $H'_{kj}$ can be constructed in a $N$ particles system physically with Ising-type interaction as given by
\begin{equation}
H_I=\sum_i(a_i\sigma_x^i+b_i\sigma_y^i+c_i\sigma_z^i)+\sum_{i<j}V_{ij}\sigma_{ij},
\end{equation}
where $\sigma_l^i=I\otimes I\cdots\otimes\sigma_l\otimes\cdots I$, $\sigma_l (l=x, y, z)$ are the Pauli matrixes that being applied to the $i$th particle, and the coefficients $a_i, b_i, c_i$ can be time- or site-dependent. $V_{ij}$ are the nearest-neighbour coupling strength and $\sigma_{ij}$ are the coupling terms which have the form $\sigma_l^i\sigma_l^j$ or $\sigma_+^i\sigma_-^j+\sigma_-^i\sigma_+^j$ , $\sigma_+^{i(j)}=(\sigma_x^{i(j)}+i\sigma_y^{i(j)})/2$ and $\sigma_-^{i(j)}=(\sigma_x^{i(j)}-i\sigma_y^{i(j)})/2$ \cite{Bernien2017,Browaeys2020,Chen2023}. By choosing suitable coefficients $a_i, b_i, c_i, V_{ij}$, Hamiltonian $H'_{kj}$ can be realized by $H_I$, i.e., given an initial state $|\Psi_0\rangle=k_{00}|00\rangle+k_{11}|11\rangle), (k_{00}^2+k_{11}^2=1)$, the braiding operation of $\{|00\rangle, |11\rangle\}$ can be realized by $H_I=\Omega_I(\sigma_x^1\sigma_x^2+\sigma_y^1\sigma_y^2)/4$, $\Omega_IT_I=\pi$. To discuss the dynamics in the degenerate eigen subspace, one may adopt suitable coupling configuration such as Hamiltonian (1) in the basis of $\{|\xi_i\rangle\}$. Therefore, the system can be extended to the Hilbert space with dimension of $2^N$ in a $N$ particles system.

\ \

\section{Conclusion}
In summary, we have proposed a scheme to realize braiding operations in a three-fold degenerate eigen subspace of a four-level system which is also the minimal physical system. The topological transition and phase coherence of the braiding factually exhibit the distinction between the quantum system and the classical system. Furthermore, we show that equivalent relationship function and phase-variation method can be combined to classify the topology of the braiding. As known, classifications of knots are NP-problem in knot theory which is still challenging \cite{Freivalds2005}. Our work may provide a feasible way to solve the problem with physical resources.

\bigskip
\begin{acknowledgments}

This work was supported by the National Natural Science Foundation of China (Grants No. 12074132, No. 12225405, No. 12247123, and No. U20A2074), the China Postdoctoral Science Foundation(Grant No. 2022M721222, No. 2023T160233).

Zhi-Wei Han and Jia-Hao Liang contribute equally to this work.
\end{acknowledgments}

\appendix*
\section{Remarks}
\subsection{Derivation of the four-level system with a three-fold degenerate eigen subspace}

To construct a four-level system with a three-fold degenerate eigen subspace directly is quite a challenging task. At the same time, it is well-known that a $N$-pod system has a degenerate eigen subspace with $N-2$ dimensions. Therefore, we may get the four-level system with a three-fold degenerate eigen subspace deduced from the five-pod system with large detuning. Under the bare state basis $\{|1\rangle, |2\rangle, |3\rangle, |4\rangle, |5\rangle\}$ and rotating-wave approximation, the interacting Hamiltonian of a five-pod system is given by
\begin{equation}
H(t)=\frac{\hbar}{2}\left(\begin{array}{cccccc}
0 & 0 & 0 & 0 & \Omega'_1 \\
0 & 0 & 0 & 0 & \Omega'_2 e^{-i \varphi} \\
0 & 0 & 0 & 0 & \Omega'_3 \\
0 & 0 & 0 & 0 & \Omega'_4 e^{-i \varphi} \\
\Omega'_1 & \Omega'_2 e^{i \varphi} & \Omega'_3 & \Omega'_4 e^{i \varphi} & -2 \Delta_5
\end{array}\right),
\end{equation}
where $\Omega'_i (i=1,2,3,4,5)$ are the Rabi frequencies and $\Delta_5$ is the single-photon detuning. According to the Schr\"{o}dinger equation and the large detuning situation ($\Delta_5\gg\Omega'_i$), we can get
\begin{equation}
\begin{split}
& \frac{d c_1}{d t}=-\frac{i \Omega'_1}{4 \Delta_5}\left(\Omega'_1 c_1+\Omega'_2 e^{i \varphi} c_2+\Omega'_3 c_3+\Omega'_4 e^{i \varphi} c_4\right), \\
& \frac{d c_2}{d t}=-\frac{i \Omega'_2}{4 \Delta_5}\left(\Omega'_1 e^{-i \varphi} c_1+\Omega'_2 c_2+\Omega'_3 e^{-i \varphi} c_3+\Omega'_4 c_4\right), \\
& \frac{d c_3}{d t}=-\frac{i \Omega'_3}{4 \Delta_5}\left(\Omega'_1 c_1+\Omega'_2 e^{i \varphi} c_2+\Omega'_3 c_3+\Omega'_4 e^{i \varphi} c_4\right), \\
& \frac{d c_4}{d t}=-\frac{i \Omega'_4}{4 \Delta_5}\left(\Omega'_1 e^{-i \varphi} c_1+\Omega'_2 c_2+\Omega'_3 e^{-i \varphi} c_3+\Omega'_4 c_4\right), \\
\end{split}
\end{equation}
where $c_i$ are the probability amplitude of bare states $|i\rangle, i=1, 2, 3, 4$, respectively.
One can find that the effective four-level system upon bare states $\{|1\rangle, |2\rangle, |3\rangle, |4\rangle\}$ is given by
\begin{equation}
H_{\mathrm{eff}}=\frac{\hbar}{2}\left(\begin{array}{cccc}
0 & \Omega_{11} e^{i\varphi} & \Omega_{13} & \Omega_{14} e^{i\varphi} \\
\Omega_{11} e^{-i\varphi}& 0 & \Omega_{13} e^{-i\varphi}& \Omega_{14} \\
\Omega_{13} & \Omega_{13} e^{i\varphi}& \Delta_{31} & \Omega_{34} e^{i\varphi} \\
\Omega_{14} e^{-i\varphi}& \Omega_{14}& \Omega_{34} e^{-i\varphi}& \Delta_{41}
\end{array}\right),
\end{equation}
with $\Omega_{11}=\Omega_1'^2/2\Delta_5, \Omega_{13}=\Omega_1' \Omega_3'/2\Delta_5, \Omega_{14}=\Omega_1' \Omega_4'/2\Delta_5, \Omega_{34}=\Omega_3' \Omega_4'/2\Delta_5, \Delta_{31}=(\Omega_3'^2-\Omega_1'^2)/2\Delta_5, \Delta_{41}=(\Omega_4'^2-\Omega_1'^2)/2\Delta_5$, and $\Omega_1'=\Omega_2'$.
One can check that Hamiltonian $H_{\textrm{eff}}$ has a three-fold degenerate eigen subspace. By replacing $\Omega_{11}$ with $\Omega_1$, $\Omega_{13}$ with $\Omega_3$, $\Omega_{14}$ with $\Omega_4$, $\Omega_{34}$ with $\Omega_2$, $\Delta_{31}$ with $\Delta_3$, and $\Delta_{41}$ with $\Delta_4$, we can recover Hamiltonian (1) from Hamiltonian (A.3).

By adopting $(N+1)$-pod configuration, the $N$-fold degenerate subspace can be deduced after the reduction. It can be found that a fully connected coupling should be realized in the reduced Hamiltonian which maybe experimentally challenging for large $N$. To solve this problem, the series-connected tripod system can be adopted \cite{Zhang2022}. It can be checked that a Hamiltonian with terms that coupling the nearest neighbour are needed after the reduction, which will improve the experimental feasibility.
\\
%%%%%%%%%%%%%%%%%%%%%%%%%%%%%%%%%%%%%%%%%%%%%%%%%
%%%%%%%%%%%%%%%%%%%%%%%%%%%%%%%%%%%%%%%%%%%%%%%%%
\begin{figure}[ptb]
\begin{center}
\includegraphics[width=5cm]{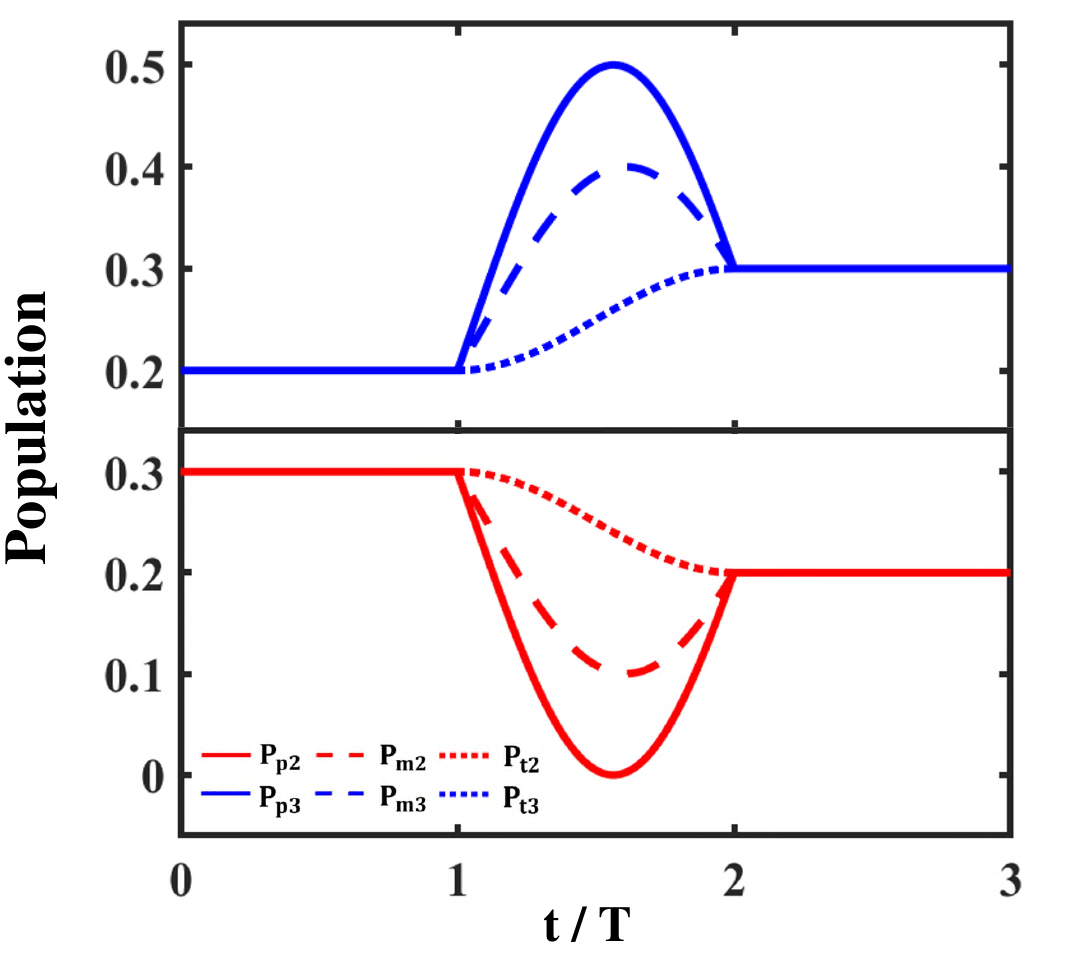}
\caption{Population dynamics of different input states where the density matrix is described by Eq. (A.4). The coherence of input states can be tilted by parameter $\eta$. Solid lines ($P_\mathrm{p2}, P_\mathrm{p3}$): $\eta=1$; Dashed lines ($P_\mathrm{m2}, P_\mathrm{m3}$): $\eta=1/2$; Dotted line ($P_\mathrm{t2}, P_\mathrm{t3}$): $\eta=0$. The red lines are the numerical results of dressed state $|\lambda_2\rangle$ and the blue lines are the ones of $|\lambda_3\rangle$.}
\end{center}
\end{figure}
%%%%%%%%%%%%%%%%%%%%%%%%%%%%%%%%%%%%%%%%%%%%%%%%%%%
\begin{figure}[ptb]
\begin{center}
\includegraphics[width=8.5cm]{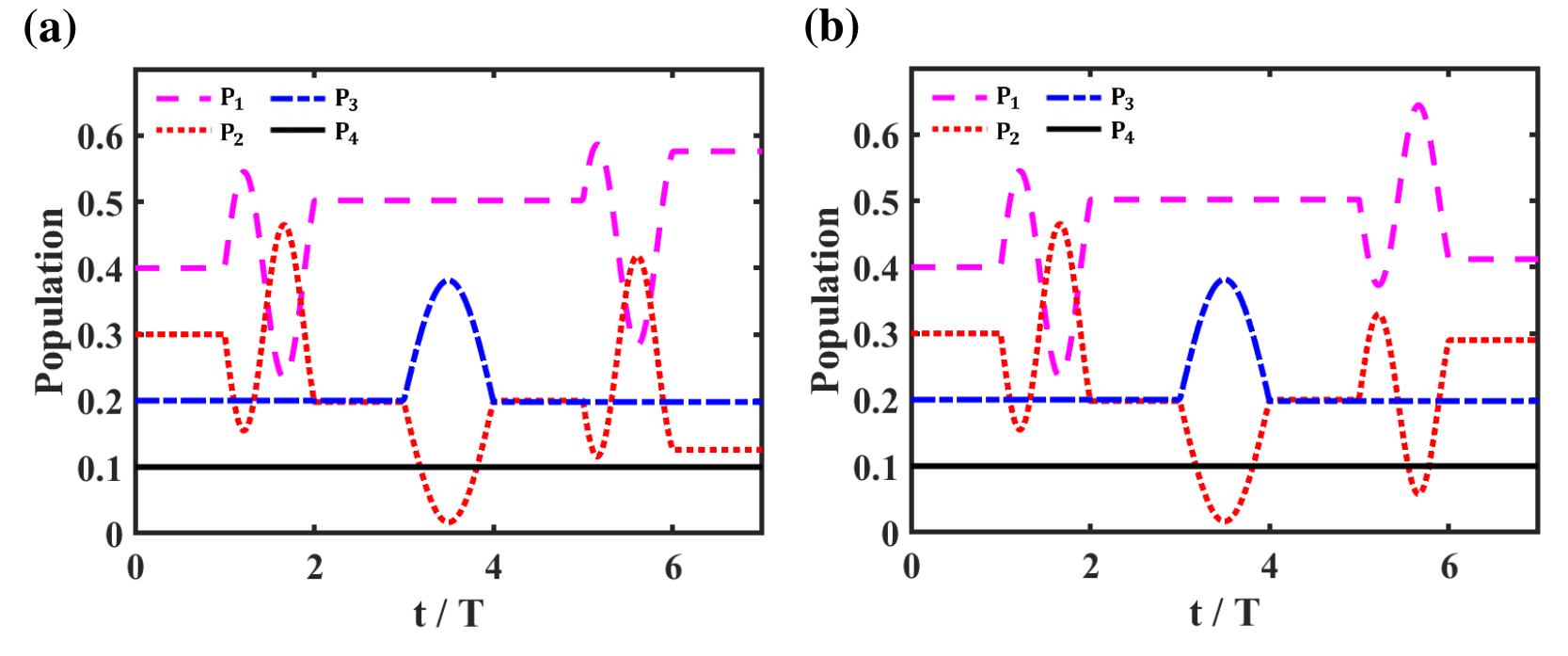}
\caption{(a) Population dynamics against braiding $\pi_{12,\mathrm{u}}\pi_{23,\mathrm{u}}\pi_{12,\mathrm{o}}$ with additional Hamiltonian $H_\mathrm{g}$ adding to all of the $\pi$ pulses. (b) Population dynamics against braiding $\pi_{12,\mathrm{o}}\pi_{23,\mathrm{u}}\pi_{12,\mathrm{o}}$ with additional Hamiltonian $H_\mathrm{g}$ adding to all of the $\pi$ pulses. Through detecting the breaking dynamics we can distinguish the braiding that with the same linking pattern but different wringing patterns. Pink dashed lines: population of $|\lambda_1\rangle$, $P_1$; Red dotted lines: population of $|\lambda_2\rangle$, $P_2$; Blue dashed-dotted lines: population of $|\lambda_3\rangle$, $P_3$; Black solid lines: population of $|\lambda_4\rangle$, $P_4$.}
\end{center}
\end{figure}
%%%%%%%%%%%%%%%%%%%%%%%%%%%%%%%%%%%%%%%%%%%%%%%%%%%

\subsection{Effect of initial mixed states on quantum braiding}

Here we discuss the distinction of braiding of different input states: pure states and mixed states, while the braiding operations are still coherent. The input states now are described by the density operator as
\begin{equation}
\rho_0=\left(\begin{array}{cccc}
\left|c_1\right|^2 & \eta c_1 c_2^* & \eta c_1 c_3^* & \eta c_1 c_4^* \\
\eta c_2 c_1^* & \left|c_2\right|^2 & \eta c_2 c_3^* & \eta c_2 c_4^* \\
\eta c_3 c_1^* & \eta c_3 c_2^* & \left|c_3\right|^2 & \eta c_3 c_4^* \\
\eta c_4 c_1^* & \eta c_4 c_2^* & \eta c_4 c_3^* & \left|c_4\right|^2
\end{array}\right).
\end{equation}
The density operator is a pure state when $\eta=1$ and is mixed state when $0\leq\eta<1$, $c_i, i=1, 2, 3, 4$ are probability amplitudes of $|i\rangle$.

In Fig. 5, the population dynamics of the braiding upon eigen states $|\lambda_{2, 3}\rangle$ are shown, where the control parameters are the same with the ones in Fig. 2 and $c_1=\sqrt{0.4}, c_2=\sqrt{0.3}, c_3=\sqrt{0.2}, c_4=\sqrt{0.1}$. We consider union operations of $\pi_{23,\mathrm{u}}$ with $\phi=\pi/2$. The density operator is computed by $\rho(t)=U_c \rho_0 U_c^{\dagger}$, $U_c=U_H(T_2)U_\pi(T)U_H(T_1)$, $U_\pi(T)=\pi_{23,\mathrm{u}}$. The red (population of $|\lambda_{2}\rangle$, $P_\mathrm{p2}$) and blue (population of $|\lambda_{3}\rangle$, $P_\mathrm{p3}$) solid lines in Fig. 5 are the numerical results of pure states with $\eta=1$ in Eq. (A4). Since the initial state and the braiding are both coherent, the dynamics of the population will be subject to the phase $\phi$ sinusoidally. The red (blue) dotted lines are the results of $P_\mathrm{t2}(P_\mathrm{t3})$ with the situation of $\eta=0$. Here the input state is classical of which the population dynamics will not change as one tilts the phase $\phi$. A common case of $\eta=1/2$ is shown by the red (blue) dashed lines in Fig. 5 where the population dynamics of $P_\mathrm{m2} (P_\mathrm{m3})$ can be subjected to the phase variation of $\phi$ but will not as perfect as the case in $\eta=1$. Therefore, a quantum braiding needs the input state and the braiding operations to be both coherent.

\subsection{Characterizing the wringing pattern with the breaking dynamics}
Taking advantage of the full controllability of the proposed system, we can test the braiding topology by breaking the braiding which can be done by adding a global  Hamiltonian to the braiding operations.

As in the case of Fig. 4(a), two braiding with the same linking pattern but different wringing patterns cannot be distinguished by the population of the output states, that is function $K$. However, through applying a specific (but not unique) Hamiltonian

\begin{equation}
H_\mathrm{g}=\Omega_0\left ( \begin{array}{cccc}
0&-1&0& 0\\
-1& 0 & 0&0\\
0&0& -1 &0\\
0&0&0 &  -1
\end{array}\right),
\end{equation}
to all of the $\pi$ pulses, the results will be different. The numerical results are shown in Fig. 6(a) and 6(b). The evolution operator of Fig. 6(a) is derived as
\begin{equation}
U_\mathrm{ag}=U_{H_\mathrm{g}-H_{12}}U_{H_\mathrm{g}-H_{23}}U_{H_\mathrm{g}+H_{12}},
\end{equation}
while the evolution operator of Fig. 6(b) is derived as
\begin{equation}
U_\mathrm{bg}=U_{H_\mathrm{g}+H_{12}}U_{H_\mathrm{g}-H_{23}}U_{H_\mathrm{g}+H_{12}},
\end{equation}
where $U_{H'}=e^{-iH'T}$. The initial states and the control parameters are the same as the ones in Fig. 2(a) and 2(b). The existence of $H_\mathrm{g}$ break the configurations of $U_\mathrm{a}$ and $U_\mathrm{b}$ since operators $U_{H_\mathrm{g}+H_{12}}(U_{H_\mathrm{g}-H_{12}})$ are no more $\pi$-pulses. The evolution will be dynamically depended on $\Omega_0T$ yet still inside the three-fold degenerate eigen subspace. In Fig. 6(a) and 6(b), one can easily find that the final populations $P_i, i=1, 2, 3,$ are different for $U_\mathrm{ag}$ and $U_\mathrm{bg}$. Consequently, we can describe the topology of braiding by function $K$ and breaking dynamics.

\subsection{Qutrit gates in degenerate eigen subspace }
$d$-ary digits ($d>2$) that encoded in multi-quantum states have emerged as an alternative way to qubit for quantum computation and quantum information science. Due to the larger Hilbert space, $d$-ary digits feature more powerful ability to do multiple control operations simultaneously and reduce the circuit complexity, simplifying the experimental setup and enhancing the algorithm efficiency \cite{Klimov2003,Wang2020}. Here we investigate the realization of geometric qutrit gates in the three-fold degenerate eigen subspace with Hamiltonian (1), i.e. , the Pauli X gate $X_3$ and the Z gate $Z_3$ with the form
\begin{equation}
X_3=\left ( \begin{array}{ccc}
0&0&1\\
1& 0 & 0\\
0&1& 0
\end{array}\right),\
Z_3=\left ( \begin{array}{ccc}
1&0&0\\
0& e^{i\phi_3} & 0\\
0&0& e^{2i\phi_3}
\end{array}\right).
\end{equation}
One can find that $X_3$ gate is the cyclic permutation of the probability amplitudes of the quantum state and can be achieved by a pulses sequence of $\pi_{12,\mathrm{o}}\pi_{23,\mathrm{o}}$. Furthermore, the $Z_3$ gate can be achieved by the sequences $\pi_{23,\mathrm{u}}(\phi_3)\pi_{23,\mathrm{o}}(0)\pi_{12,\mathrm{u}}(\phi_3)\pi_{12,\mathrm{o}}(0)$ where $0$ or $\phi_3$ in $\phi_{kj,\mathrm{o(u)}}$ can be determined by the $\phi_{kj}$ in Eq. (5) and (6). The operation $\pi_{kj,\mathrm{o(u)}}$ can be realized geometrically and thus the $X_3, Z_3$ gates \cite{Sj2012,Abdumalikov,Zu2014,Toyoda2013,Zhu2002,Xue201802,Xue201801,Xue2021,Xue2023,Du2017}. The present control can be also generalized to the case of $d$-ary digits with n-fold degenerate eigen subspace. Since the eigen states are connected to the bare states with Eqn. (7), therefore, the proposed gates can be applied to the bare states after a unitary transformation.

%%%%%%%%%%%%%%%%%%%%%%%%%%%%%%%%

\end{document}